# Application of Liquid Rank Reputation System for Twitter Trend Analysis on Bitcoin

Abhishek Saxena
*Mathematics and Mechanics Department*
*Novosibirsk State University*
Novosibirsk, Russia
abhishek.saxena.mnnit@gmail.com

Anton Kolonin
*Mathematics and Mechanics Department*
*Novosibirsk State University*
Novosibirsk, Russia
akolonin@gmail.com

*Abstract*—Analyzing social media trends can create a win-win situation for both creators and consumers. Creators can receive fair compensation, while consumers gain access to engaging, relevant, and personalized content. This paper proposes a new model for analyzing Bitcoin trends on Twitter by incorporating a 'liquid democracy' approach based on user reputation. This system aims to identify the most impactful trends and their influence on Bitcoin prices and trading volume. It uses a Twitter sentiment analysis model based on a reputation rating system to determine the impact on Bitcoin price change and traded volume. In addition, the reputation model considers the users' higher-order friends on the social network (the initial Twitter input channels in our case study) to improve the accuracy and diversity of the reputation results. We analyze Bitcoin-related news on Twitter to understand how trends and user sentiment, measured through our Liquid Rank Reputation System, affect Bitcoin price fluctuations and trading activity within the studied time frame. This reputation model can also be used as an additional layer in other trend and sentiment analysis models. The paper proposes the implementation, challenges, and future scope of the liquid rank reputation model.

## I. Introduction

The biggest challenge in analyzing social media trends lies in identifying genuine influence, particularly regarding the impact of figures like Matthew on social commerce [1]. Another significant concern is the manipulation of trends through tactics like bot armies or fabricated data, as evidenced by the spread of COVID-19 misinformation via Twitter bots [2]. This raises crucial questions about the validity of trend indicators and the need for robust validation mechanisms.

To address these issues, a reputation system "unpublished" [3] has been proposed, leveraging the principles of "liquid democracy" to prioritize trustworthy voices and mitigate the influence of manipulative actors. This system aims to create a more reliable and equitable landscape for understanding and analyzing social media trends.

One of the solution is to use mechanism design approaches in a reputation algorithm to detect the impact of a trend based on a particular person's higher-order friends or peer-to-peer recommendations [4].

This paper explores how the "liquid democracy" model can be implemented for analyzing sentiment in Bitcoin-related Twitter news using a "liquid reputation system" "unpublished" [3]. We achieve this by analyzing a dataset of Bitcoin news tweets, aiming to understand the impact of this reputation system on both Bitcoin price fluctuations and trading volume. The system assigns varying weights to users based on their reputation, influencing the overall sentiment analysis of tweets within a specific time frame.

This implementation operates on the reputation of their "higher-ranking insiders" or friends on Twitter. This is achieved by selecting relevant channels with validated reputation scores from the Twitter sentiment analysis data. By incorporating this "liquidity rating," we observe improved correlations between sentiment analysis results and actual Bitcoin price changes and trading volume. Similar approaches utilizing higher-order friend networks for content recommendation have been explored in previous research [5].

### A. Why Liquid Rank Reputation System?

The goal of this system is to create a trustworthy and open environment where information is managed fairly. In decentralized networks where every participant interacts directly [6], a reliable reputation system is crucial. The reputation system functions by calculating a score for each participant based on their past actions and contributions across various platforms. This can include engagement on social media, communication channels, financial transactions, or any other platform where their behavior can be tracked over time. By leveraging data from diverse sources, the system paints a holistic picture of each participant's trustworthiness and reliability, ultimately relying on the collective involvement of all network members as verifiers.

Section II presents the Methodology, Section III Results. Section IV Future work and scope, and Section V Conclusion.

## II. Methodology

### A. Implementation of Liquid Democracy in the Twitter Bitcoin Sentiment Analysis Model using the Liquid Rank Reputation algorithm.

This reputation system design draws inspiration from previous research on "social capital" and "karma" models "unpublished" [3]. By applying the principles of "liquid democracy" to a dataset of Bitcoin-related tweets scraped from

Twitter, it aims to identify correlations between the sentiment of trends and actual changes in Bitcoin price and trading volume within a specific time frame.

Two kinds of metrics were derived from the online social media data: public posts from about 80 channels on Twitter relevant to the cryptocurrency market on Bitcoin for 1 year and six months starting in June 2021. Hence, initial channels serving as inputs can be considered as higher-order friends on a social platform. First, it was the conventional sentiment scores, computed as described below. Second, it was the "cognitive behavioral schemata" (CBS) patterns. The overall volume of media content exceeded 397,117 posts across 48824 channels. The dataset contains all 21 metrics, including CBS, sentiment scores, and tweet descriptions, along with time stamp, word count, and item count.

Apart from that, BTC price and volume data was collected for price and volume trends from June 2021 to December 2022. This dataset contains BTC price open, close, volume, high/low BTC/USD price, and time stamp.

This paper analyzes a Twitter dataset containing Bitcoin news, previously processed for sentiment analysis. Applying a liquid rank reputation system, we identify the user with the highest reputation score, essentially acting as the "reputation leader" within this dataset. This leader could be considered a potential starting point for a content recommendation system. Fig. 1 visualizes the ranking of users based on this liquid rank reputation system.

The paper makes the use case simple, also considering the restraints of the dataset (not getting all the metrics like followers, likes, etc.). A simple computational model of reputation [2] is calculated as

$$R_j = \sum (R_i \times V_{ijt}) \quad (1)$$

where $V_{ijt}$ represents a positive implicit rating. It signifies how often channel j (a Twitter channel being evaluated) is mentioned by channel i (the one providing the rating) within a specific timeframe (t). The reputation score is calculated for individual channels, but each channel is considered to act as a multi-agent social network representative for its associated user accounts. This indirect rating occurs through the channels selected as initial data sources (considered "higher-order friends") during dataset collection for the given time frame.

The algorithm applies normalization after each round of calculating reputation scores, then repeats the process until the changes in reputation become negligible (less than 0.0001). Reputation score of each unique user gets multiplied with the related tweet sentiment metrics, CBS, word count, and item count as

Reputation_'metric' = metric x Aggregated Reputation Score (2)

The dataset has been aggregated day-wise with mean values after all the channels have been mapped with their respective reputation scores. Along with that, we have also aggregated BTC-USD data day-wise to correlate the price and volume trend with reputation metrics between June 2021 and December 2022.

Finally, for qualitative analysis, the correlation is calculated using Pearson correlation analysis for:

- Metrics vs. Price Change and Volume
- Reputation Metrics vs. Price Change and Volume

where 21 metrics of the sentiment analysis and CBS (namely sen, pos, neg, con, catastrophizing, dichotoreasoning, disqualpositive, emotionreasoning, fortunetelling, labeling, magnification, mentalfiltering, mindreading, overgeneralizing, personalizing, shouldment, exclusivereasoning, negativereasoning, and mentalfilteringplus) along with wordcnt and itemcnt (all column names from the dataset which have evident meanings as names) versus Bitcoin price change and volume traded. Pearson correlation for the same metrics updated with the liquid rank reputation system versus Bitcoin price change and volume traded.

All the metrics, including reputation metrics, have been aggregated day-wise to have daily results and predictions. Hence, for the impact of the above metrics, we have further calculated Pearson Correlation:

- With -1 Day Lagged BTC-USD Data and
- Without Lag BTC-USD Data

We saw the results have fine Pearson correlation scores but not any major impact can be withdrawn from them. So, we further study the process of finding what we called a synthetic additive cause indicator (SACI), relying on the whole scope of source metrics being treated as hypothetical causes. The probabilistic logic treats addition as logical disjunction and multiplication as logical conjunction. In this work, we were exploring only the disjunctive version of it, so the assembly of the integrative SACI involved the addition of the perspective causes in order to maximize the correlation with the effect at a particular target shift or lag.

The temporal causation study was evaluated with and without lag (1 day) for all metrics and reputation metrics, computing the mutual Pearson correlation between each of the potential causes and the price difference and retaining the "correlation weights" of the computed value $P(l,m)$ for lag l and metric m. Also, the channels c were optionally weighted with the "reputation weight" as $R(c)$ according to the mentions in Twitter feed present at such time intervals, as shown in Fig. 1, below. Then, for lag l, the compound SACI metric time series:

$$Y(l,d) = \sum X(c,m,d) \times P(l,c,m) \times R(c) \quad (3)$$

for day d has been built from the original raw metrics or reputation metrics $X(c,m,d)$ in the case of reputation-scored metrics. The compound SACI metric building process was implemented starting from channels with the highest $R(c)$ and $P(l,c,m)$ adding ingredients up to $Y(l,d)$ as the correlation between the target price difference, volume function, and related metrics. Fig. 4 and 5 give the final SACI (additive) for price change Pearson correlation and volume Pearson correlation of all the metrics and reputation metrics with and without a lag of 1-day.

III. RESULTS

A. *Pearson Correlations for Metrics and Reputation Metrics with and without lag*

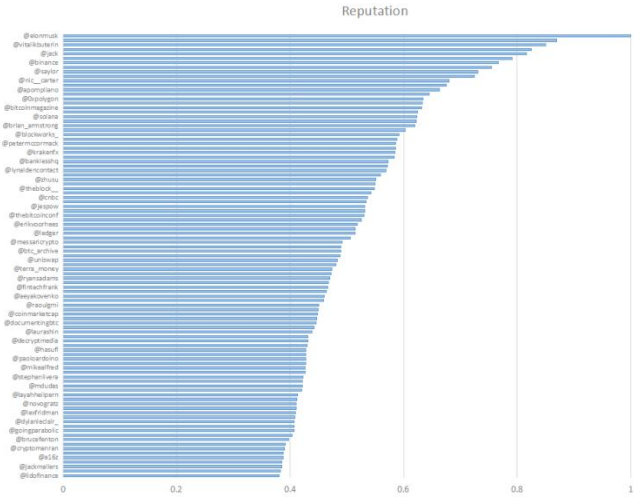

Fig. 1. Reputation Weights R(c) (scores) of the Top 100 channels

Fig. 1 shows the top 100 users or channels in BTC Twitter dataset after liquid rank reputation system implementation. The list of rankings is based on reputation scoring from equation (1). These are the impacting channels in the dataset, concerning sentiments and the CBS of the dataset.

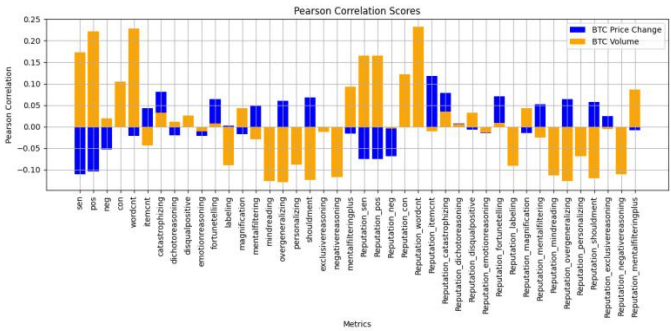

Fig. 2. Pearson Correlations for Metrics and Reputation Metrics without lag

Fig. 2 shows the impact of the reputation system on the sentiment analysis model without lag. The main impact can be seen in a few metrics, like Reputation_wordcnt and Reputation_con on volume traded in BTC. We see that the increment correlation score is of order 2-3% in the above metrics using the reputation system.

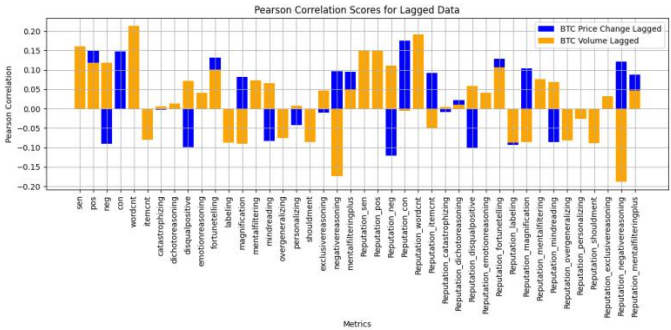

Fig. 3. Pearson Correlations for Metrics and Reputation Metrics with 1 Day lag

Fig. 3 shows the impact of the reputation system on the sentiment analysis model with 1 day lag. The impact is seen in metrics Reputation_neg, Reputation_con, Reputation_itemcnt, Reputation_negativereasoning and Reputation_magnification on the price change of BTC. There is a major impact of Reputation_negativereasoning on volume traded in BTC. Also we see that lagged data have better correlation in the above-mentioned price change metrics with reputation systems, as in SACI, where we can see ~10% of an increment in correlation.

### B. Synthetic Additive Cause Indicator (SACI) for Metrics and Reputation Metrics with and without lag

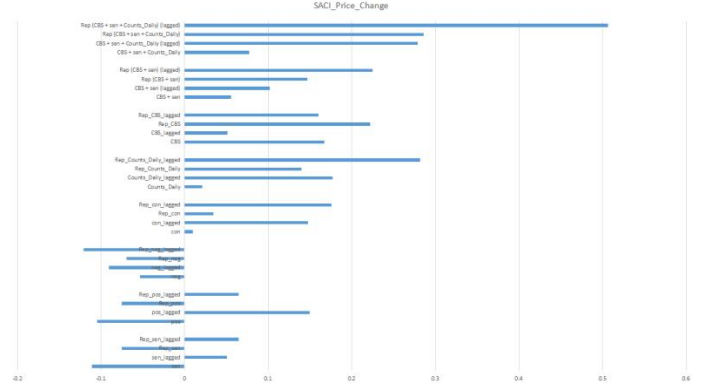

Fig. 4. SACI for Price Change Pearson Correlation

The SACI bar chart for price change metrics of Bitcoin above has added metrics for qualitative analysis using SACI, including sen (total sentiment score), sen lagged, pos (positive sentiment score), pos lagged, neg (negative sentiment score), neg lagged, con (contradictive sentiment score), con lagged, Counts_Daily (wordcnt+itemcnt), Counts_lagged, CBS (cognitive behavioral schemata), CBS lagged, CBS+sen, CBS+sen lagged, CBS+sen+Counts_Daily, CBS+sen+Counts_Daily lagged.

Now, we can easily see the impact of the reputation system on the Bitcoin Twitter analysis. In Fig. 4 we have additive indicator Rep (CBS+sen+Counts_Daily) (lagged), Rep (CBS+sen) (lagged) Rep_Counts_Daily_lagged and Rep_con_lagged giving us the overall score of ~0.5, ~0.21, ~0.28 and ~0.18 respectively. These scores are more than 50% higher than the respective scores without the application of the reputation system in SACI metrics. Another conclusion can be drawn that the 1-day lag has more impact on the price change of Bitcoin than the volume traded in Bitcoin.

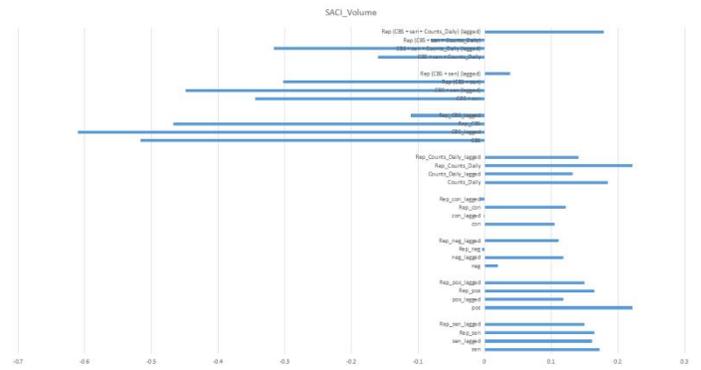

Fig. 5. SACI for Volume Pearson Correlation

The SACI bar chart for volume metrics of Bitcoin above has added metrics for qualitative analysis using SACI, including sen (total sentiment score), sen lagged, pos (positive sentiment score), pos lagged, neg (negative sentiment score), neg lagged, con (contradictive sentiment score), con lagged, Counts_Daily (wordcnt+itemcnt), Counts_lagged, CBS (cognitive behavioral schemata), CBS lagged, CBS+sen, CBS+sen lagged, CBS+sen+Counts_Daily, CBS+sen+Counts_Daily lagged.

Here, we can easily see the impact of the reputation system on the Bitcoin Twitter analysis. In Fig. 5 We have additive indicators Rep (CBS+sen+Counts_Daily) (lagged), Rep (CBS+sen) (lagged), Rep_Counts_Daily and Rep_con_lagged giving us an overall score of ~0.18, ~0.04, ~0.21, and ~0.11, respectively. These scores are 5-50% higher than the respective scores without the application of the reputation system in SACI metrics. The biggest impact we can see is in Rep (CBS+sen+Counts_Daily) (lagged) and Rep (CBS+sen) (lagged), where the reputation system was able to reverse the relation to positive magnitude in favor of correlation. Another conclusion can be drawn that the lag has less impact on the volume traded of Bitcoin than the price change of Bitcoin.

*C. Cumulative Results of the Application of Liquid Rank Reputation System on Bitcoin Twitter Sentiment Analysis.*

- Reputation system has fine-tuned the correlation results for better, on Bitcoin price change and volume traded.
- Correlation of lagged data is higher than without lag data for Bitcoin price change, and for without lag data, there is a greater impact on the volume traded of Bitcoin.
- The best correlation impact is shown in metrics, namely Reputation_wordcnt and Reputation_con on the volume traded of Bitcoin without lag condition (of order 2-3%). Reputation_neg, Reputation_con, Reputation_itemcnt, Reputation_negativereasoning and Reputation_magnification on price change of Bitcoin in a 1-day lagged condition (of order ~10%).
- SACI for volume traded vs. metrics and reputation metrics have shown higher scores (highest being > 0.5).
- SACI cumulative effect of Rep (CBS+sen+Counts_Daily) (lagged) and Rep_Counts_Daily_lagged are the best indicators of impact of Twitter trend on the prediction of Bitcoin price change in a 1-day lagged condition (showing the magnitude of the ~0.5 correlation).

## IV. FUTURE WORK AND SCOPE

By applying Liquid Ranking to multiple attributes, we can achieve two key benefits. Firstly reputation scores become more nuanced and relevant to individual users, leading to a more personalized experience on the platform. Secondly this approach helps new users and channels with lower initial influence gain visibility and recognition based on their contributions. Their comments, likes, and sentiment towards specific topics can contribute meaningfully to their reputation score, making them more relevant within the platform.

Liquid ranking can be tailored to individual users by incorporating their social media network assessing the genuine reputation of other users within their network.

Each element used in the Liquid Ranking process can be considered a time-stamped transaction, representing an interaction or activity. This allows for dynamic reputation scores that continuously update based on new data and machine learning techniques.

The strength of this reputation scoring system lies in its adaptability. It can be tailored to various environments and purposes, ranging from open social platforms and trend authorization systems to more closed, centralized settings.

In addition, it can be based on the required computing performance at certain periods, such as hourly, daily, weekly or monthly levels. (We can add more levels for features like sharing, promoting, etc. using the reputation depth model in liquid democracy).

## V. CONCLUSION

This paper shows the implementation of a liquid rank reputation system on the Twitter Bitcoin sentiment analysis dataset to find the best metrics for achieving better correlation with Bitcoin price change and volume traded.

To ensure a clear and feasible analysis within the available time and data constraints, we focused on the most readily accessible metric: mentions of specific channels within a set time frame. To assess the relationship between these mentions and the desired outcomes (Bitcoin price change and trading volume), we employed two qualitative analysis methods: Pearson correlation and the Synthetic Additive Cause Indicator (SACI).

Our analysis, utilizing the liquid rank reputation system on various qualitative metrics, revealed a significant improvement in their relevance and impact on understanding Bitcoin price behavior and trading volume within the time frame and dataset limitations. This suggests that the liquid rank reputation system, when applied to diverse metrics, holds promise for developing stronger and more nuanced correlation or prediction models.

The results of this study can help provide a better prediction metric model to users interested in Bitcoin behavior.

The liquid rank reputation system holds significant commercial potential due to its versatility and adaptability. Integrating liquid rank as an additional layer to existing prediction models can create personalized and dynamic predictors. By acknowledging reputation scores, the system encourages users to refine their contributions, leading to stronger correlations.